\documentclass{appolb}
\usepackage{graphicx}
\usepackage{float}
\usepackage[symbol]{footmisc}
\usepackage{amsbsy}
\usepackage{amsmath,amssymb,amsthm,bbm}
\usepackage{hyperref}
\usepackage{cite}
\usepackage{amsmath,amssymb,amsthm,bbm,color}
\usepackage{graphicx}
\usepackage{epsfig}
\usepackage{euscript}
\usepackage{pslatex}

\newcommand{\KK}{${\cal KK}$}
\def\rQCED{{\rm QCED}}
\newcommand{\qb}{{\bar{q}}}
\newcommand{\sfac}{\mathfrak{s}}
\begin{document}
\title{YFS Exponentiation -- Gate to Precision Accelerator Physics Experiments%
\thanks{Presented at 2024 Krakow Epiphany Conference in memory of Prof. Stanislaw Jadach}%
}
\author{B.F.L. Ward
\address{Baylor University, Waco, TX, USA}
\\[3mm]
}
\maketitle
\begin{abstract}
We present the history and current status of the YFS Monte Carlo approach to precision theory for accelerator physics experiments. Key contributions of Prof. Stanislaw Jadach are highlighted.
\end{abstract}
\centerline{BU-HEPP-24-01, Mar. 2024}
\baselineskip=11pt
\section{In Memoriam}
\begin{figure}[ht]
\begin{center}
\setlength{\unitlength}{1in}
\begin{picture}(6,4.5)(0,0)
\put(0.25,0){\includegraphics[width=4in]{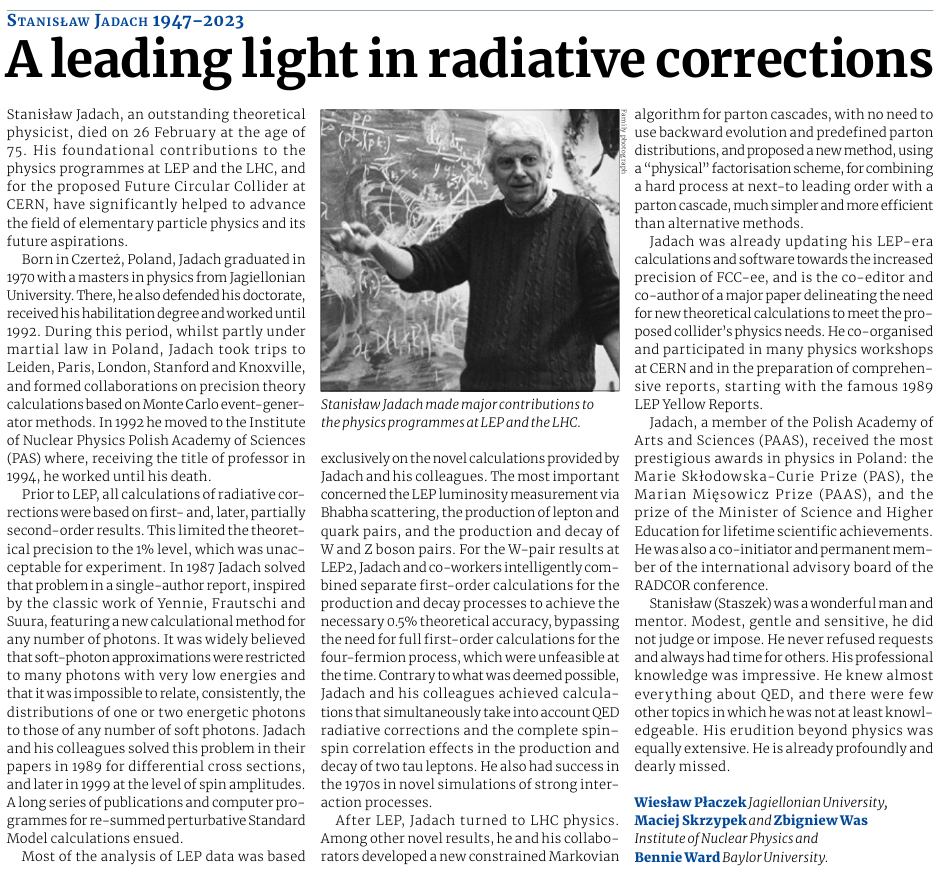}}
\end{picture}
\end{center}
\vspace{-5mm}
\caption{{\it CERN Courier} obituary for the late Prof. Stanislaw Jadach.}
\label{fig1}
\end{figure}

Sadly, my close friend and collaborator, Prof. Stanislaw Jadach, passed away suddenly on February 26, 2023. His {\it CERN Courier}\footnote{{\it CERN Courier}, May/June 2023 issue, p.59.} obituary is reproduced here in Fig.~\ref{fig1}. 
In one of his many service efforts, he was a pioneering member of the RADCOR International Advisory Board and he was the Chair of the Local Organizing Committee when the Institute of Nuclear Physics hosted the 1996 RADCOR in Krakow, Poland. With his many outstanding scientific contributions, he helped to keep our field alive. We all miss him dearly. My contribution is dedicated in memoriam to him.\par  
\section{Introduction}
To begin, the key question that we want to address is, " How did we get started on our YFS (Yennie-Frautschi-Suura~\cite{yfs:1961}) journey?"  When a collaboration has lasted for 37 years, it is appropriate to spend some time
on its origin.\par
The main event in this connection was the 1986 Mark II Radiative Corrections Meeting organized by Professor Gary Feldman at SLAC. The meeting was organized for preparation for 'precision Z physics'. The Mark II experiment did not turn on until 1989 and 
observed $\sim 750\; Z$'s\cite{burchat:1990,coupal:1990,jacobsen:1991,wagner:1992}. But, Staszek and I met at this meeting.\par
At that time, there was a 'no-go' type belief that the Jackson-Scharre~\cite{jackson-sharre:1975} naive exponentiation-based methods were the best resummation
that one could do. We started to discuss whether the methods of Yennie, Frautschi, and Suura could do better. For, the latter methods worked at the level of the amplitudes. A major question was whether or not a Monte Carlo event generator could realize all that? If so, would renormalization group improvement be possible? 
\par
The discussion was aided by our participation in the 27th Cracow School of Theoretical Physics in Zakopane, PL.  We made many long walks in the mountains discussing how we could realize the YFS theory by MC methods using the approach already written down in Staszek's MPI-Munich preprint~\cite{yfs-mpi:1987},
"Yennie-Frautschi-Suura soft photons in Monte Carlo event generators". We presented~\cite{bflward:1988-cs} the renormalization group improvement of the approach in the School.\par
As a 'proof of principle' result, we published the paper "Exponentiation of Soft Photons in Monte Carlo event generators: The Case of
Bonneau and Martin cross-section"\cite{yfs1:1988} in which the MC YFS1 with exact $\cal{O}(\alpha)|_{QED}$ and YFS resummation of soft photons to all orders in $\alpha$ was presented for initial state radiation (ISR). This MC was followed by the application of the approach to the luminosity problem 
in the paper "Multiphoton Monte Carlo event generator for Bhabha scattering at small angles"\cite{jadach:1989bhl1} in which the MC BHLUMI was presented with for low angle Bhabha scattering with exact $\cal{O}(\alpha)|_{QED}$ and YFS resummation of soft photons to all orders in $\alpha$. These two applications showed that our new YFS MC approach was amenable to processes with hard momentum transfers in both the $s$ and $t$  channels. \par
The latter two MC's were followed by our development of the MC YFS2 in the paper "YFS2: The Second Order Monte Carlo for Fermion Pair
Production at LEP/SLC with the Initial State Radiation of Two Hard and
Multiple Soft Photons"\cite{yfs2:1990} which features YFS resummation of ISR with the exact result for two hard real photons. From YFS2, we, with our collaborators, were led naturally to the development of the YFS MC's KORALZ 3.8~\cite{Jadach:1991yv} and BHLUMI 2.01~\cite{bhlumi2:1992} which realized the 
state-of-the-art precisions for fermion pair production and luminosity processes respectively which were 0.2\% and 0.25\% respectively at the time of their releases. The extension of the methods 
in YFS2 to the final state radiative effects was accomplished in Ref.~\cite{yfs3-pl:1992}. This latter
step enabled the precisions of the KORALZ and BHLUMI MC's to be improved in the versions KORALZ 4.0~\cite{Jadach:1993yv} and BHLUMI 4.04~\cite{bhlumi4:1996} to the state-of-the-art results 0.1\% and 0.11\%, respectively. The latter precision was subsequently lowered~\cite{bhlumi-precision:1998} to 0.061\% (0.054\%) if one does not (does) account for the soft pairs correction~\cite{BW11,BW12,ON1,ON2}.
\par
Building on the successes of BHLUMI and KORALZ, we and our collaborators developed further YFS MC realizations for wide angle Bhabha scattering (BHWIDE)~\cite{bhwide:1997}, the inclusion of soft pairs in the YFS resummation in low angle Bhabha scattering (BHLUMI 2.30)~\cite{BW12},$Z$-pair production (YFSZZ)~\cite{yfszz:1997} all four-fermion final states in $e^+\; e^-$ collisions (KORALW 1.42)~\cite{koralw:1998}, and $W$-pair production (YFSWW3)~\cite{yfsww3:2001}. The concurrent combination of KORALW and YFSWW3 was
developed in Ref.~\cite{kandy-2001}. These MC's played an essential role in allowing the LEP1 and LEP2 data to be analyzed with sufficient precision~\cite{bflw-smat50} to prove the correctness of the 't Hooft-Veltman renormalization program~\cite{BW1a,BW1a-2} for the $SU_{2L}\times U_1$  EW theory~\cite{SM1,SM3,SM4} as well as to help to verify the predicted running of the strong coupling constant by the Gross-Wilczek-Politzer~\cite{BW1b,BW1c} $SU_3^c$ theory of the strong interaction.\par
With the anticipation of the need to control, on an event-by-event basis, the resummation of quantum interference effects, we introduced in Ref.~\cite{ceex1:1999} the coherent generalization of the YFS theory in which IR singularities for both real and virtual photons are isolated and
resummed at the level of the quantum amplitudes. The new coherent exclusive exponentiation (CEEX) theory was then used to develop the MC's \KK{MC}~\cite{kkcpc:1999,Jadach:2000ir}, \KK{MC} 4.22~\cite{Jadach:2013aha}, \KK{MC-ee}~\cite{Arbuzov:2020coe}, \KK{MC-hh}~\cite{kkmchh,kkmchh1}, and  
\KK{MC-ee} (C++)~\cite{Jadach:2022vf}. The original \KK{MC} carried only incoming $e^+e^-$ states. \KK{MC 4.22} extended this
to all incoming $f\bar{f}$ where f is a Standard Theory~\cite{djg-smat50} fermion. \KK{MC-hh} then carries incoming hadron-hadron states such as as incoming $pp$ states.\par
\begin{figure}[h]
\begin{center}
\setlength{\unitlength}{1in}
\begin{picture}(6.5,3.0)(0,0)
\put(0,0.5){\includegraphics[width=2.5in,height=2.6in]{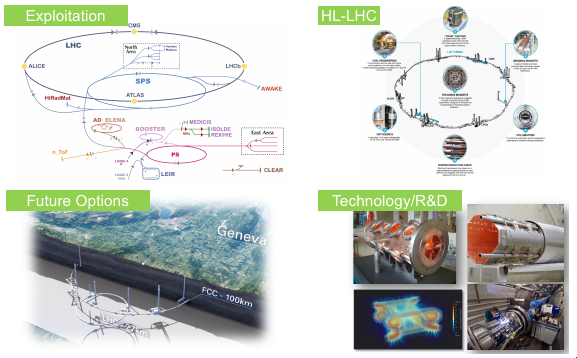}}
\put(2.6,0.5){\includegraphics[width=2.5in,height=2.6in]{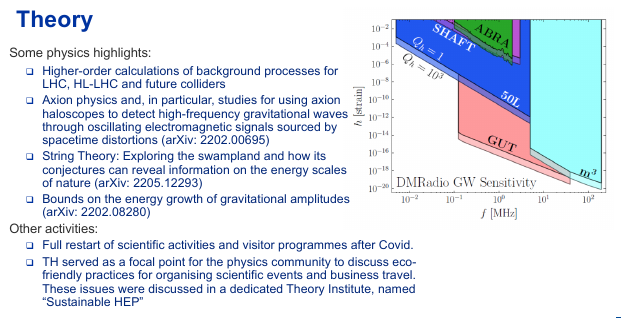}}
\put(1.5,0.25){(a)\hspace{2.8in}(b)}
\end{picture}
\end{center}
\vspace{-10mm}
\caption{\baselineskip=11pt Excerpts from Ref.~\cite{fabiola-1-10-23} on the state of CERN: (a), Future Options and R\& D; (b), Theory highlights.}
\label{fig2}
\end{figure}

The YFS MC's described above have had state-of-the-art use at SLC, LEP1 and LEP2, BaBar, BELLE, BES, the $\Phi$-Factory, and LHC for the analysis of cutting edge collider data.
Some of them are, have been, and/or will be in use also for the preparation of the physics cases for the projects TESLA, ILC, CLIC, FCC, SSC-RESTART, CEPC, CPPC, ... . What we can say is that the future of precision theory is dictated by future accelerators -- FCC, CLIC, ILC, CEPC, CPPC,  SSC-RESTART, ... . Using FCC as an example, factors of improvement 
from $\sim 5$  to $\sim 100$ are needed from theory. As we see in Fig.~\ref{fig2} with excerpts from Ref.~\cite{fabiola-1-10-23}, at world-leading laboratories such as CERN the need for precision theory for the success of future collider physics programs is recognized. The figure shows the future options for CERN featuring the FCC with the important role of higher-order calculations for its background processes as a theory highlight -- we hope the funding agencies appreciate the implied connection. Resummation is key to such calculations in many cases. In what ensues, we discuss amplitude-based resummation following
the YFS MC methodology made possible by Staszek’s seminal contributions.\par
We point out that, in contrast to methods such as the collinear factorization resummation method recently done to subleading log level in Refs.~\cite{frixione-2019,bertone-2019,frixione-2021,bertone-2022} in which various degrees of freedom are integrated out engendering an intrinsic uncertainty, the YFS approach has no limit in principle to its precision~\cite{bardin-zplep1:1989} as long as one calculates the corresponding hard radiation residuals to the desired order in the respective coupling. Here, after giving a brief recapitulation of the exact amplitude resummation theory in the next Section, we present in Section 4 illustrative results in 
both collider physics and quantum gravity to capture the expanse of the attendant methods.
In Section 5 we discuss improving the collinear limit of YFS theory -- one of our last works with Staszek. Section 6 contains our summary.\par
\section{Recapitulation of Exact Amplitude-Based Resummation Theory}
In this Section, we include a brief recapitulation of exact amplitude-based resummation theory. The starting point is the master formula that exhibits the theory which reads as follows:
{\small
\begin{eqnarray}
&d\bar\sigma_{\rm res} = e^{\rm SUM_{IR}(QCED)}
   \sum_{{n,m}=0}^\infty\frac{1}{n!m!}\int\prod_{j_1=1}^n\frac{d^3k_{j_1}}{k_{j_1}} \cr
&\prod_{j_2=1}^m\frac{d^3{k'}_{j_2}}{{k'}_{j_2}}
\int\frac{d^4y}{(2\pi)^4}e^{iy\cdot(p_1+q_1-p_2-q_2-\sum k_{j_1}-\sum {k'}_{j_2})+
D_\rQCED} \cr
&{\tilde{\bar\beta}_{n,m}(k_1,\ldots,k_n;k'_1,\ldots,k'_m)}\frac{d^3p_2}{p_2^{\,0}}\frac{d^3q_2}{q_2^{\,0}},
\label{subp15b}
\end{eqnarray}}
where the {\em new}\footnote{The {\em non-Abelian} nature of QCD requires a new treatment of the corresponding part of the IR limit~\cite{Gatheral:1983} so that we usually include in ${\rm SUM_{IR}(QCED)}$ only the leading term from the QCD exponent in Ref.~\cite{Gatheral:1983} -- the remainder is included in the residuals $\tilde{\bar\beta}_{n,m}$ .}(YFS-style) residuals   
{$\tilde{\bar\beta}_{n,m}(k_1,\ldots,k_n;k'_1,\ldots,k'_m)$} have {$n$} hard gluons and {$m$} hard photons. We refer the reader to Refs.~\cite{mcnlo-hwiri,mcnlo-hwiri1} for the definitions of the new residuals and the infrared functions ${\rm SUM_{IR}(QCED)}$ and ${ D_\rQCED}$.  Parton shower/ME matching engenders the replacements {$\tilde{\bar\beta}_{n,m}\rightarrow \hat{\tilde{\bar\beta}}_{n,m}$}, as explained in Ref.~\cite{mcnlo-hwiri,mcnlo-hwiri1}. 
Using the basic formula{\small
\begin{equation}
{d\sigma} =\sum_{i,j}\int dx_1dx_2{F_i(x_1)F_j(x_2)} d\hat\sigma_{\rm res}(x_1x_2s),
\label{bscfrla}
\end{equation}}
the latter replacements allow us to connect with MC@NLO~\cite{mcnlo,mcnlo1}.
\par
Eq.(\ref{subp15b}) has been used to obtain new results in precision LHC and FCC physics. In a new approach to quantum gravity~\cite{ward:2013dkunv,ijmpa2018}, we have extended the latter equation to general relativity. In the next Section, we discuss such new results and their attendant new perspectives with an eye to their role as a gate to precision collider physics.\par
\section{Gate to Precision Collider Physics: LHC, FCC, CPEC, CPPC, ILC, CLIC}
\begin{figure}[h!]
\begin{center}
\setlength{\unitlength}{1in}
\begin{picture}(6,2.0)(0,0)
\put(0,0){\includegraphics[width=5in]{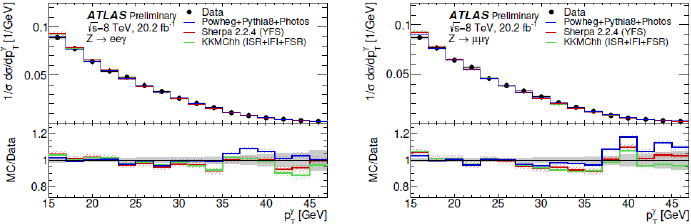}}
\end{picture}
\end{center}
\vspace{-5mm}
\caption{ATLAS analysis of $Z/\gamma$ production at $8$ TeV.}
\label{fig3}
\end{figure}

The MC event generator \KK{MC}-hh~\cite{kkmchh1-sh} carries a representation of eq.(\ref{subp15b}) which illustrates how the YFS methodology continues to be a gate to precision collider physics by allowing a new perspective on the expectations for precision physics for the Standard Theory
EW interactions at HL-LHC. This is shown by the plots in Fig.~\ref{fig3} which are taken from the ATLAS analysis~\cite{atlas-epjc-zg:2024} of $Z\gamma$ production at 8 TeV.
The data for the $\gamma\; p_T$ spectrum are compared to the Powheg-Pythia8-Photos~\cite{powheg-org,powheg-orga,powheg1,powhega,Sjostrand:2007gs-sh,Golonka:2006tw}, Sherpa2.2.4(YFS)~\cite{sherpa,sherpa-2.2}, and \KK{MC}-hh predictions. At the current level of the uncertainties in the data 
all three predictions are in reasonable agreement with the data. At HL-LHC, we expect ~10 times the current statistics so that a precision test against the theories will obtain.\par
\begin{figure}[h!]
\begin{center}
\setlength{\unitlength}{1in}
\begin{picture}(6,2.0)(0,0)
\put(0,0){\includegraphics[width=5in]{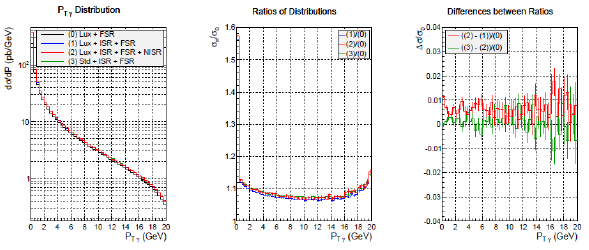}}
\end{picture}
\end{center}
\vspace{-5mm}
\caption{The distribution for $P_{T_\gamma}$ of the photon for which it is greatest for events with at least one photon and each lepton having $ p_{T\ell}> 25 GeV, \eta_\ell< 2.5$ calculated with (0) FSR only (black). (1) FSR + ISR (blue). and (2) FSR + ISR with NISR (red) for NNPDF3.1-LuxQED NLO PDFs. For comparison, (3) shows FSR + ISR with ordinary NNPDF3.1 NLO PDFs (green). The center graph shows
ISR on/off ratios (1)/(0) (blue),(2)/(0) (red) and (3)/(0) (green). The right-hand graph shows the fractional differences ((1)- (2))/(0) in red and ((2)- (3))/(0) in green.}
\label{fig4}
\end{figure}

 
An important issue is the effect of QED contamination in non-QED PDFs. In another illustration of the gateway aspect of the YFS methodology for precision accelerator physics, to  resolve this issue~\cite{jad-yost-afb,ichep2022-say,sjetaltoappear} we use Negative ISR (NISR) evolution to address the size of this contamination directly. Using a standard notation for PDFs and cross sections, the cross section representation is
\begin{equation}
\begin{split}
\sigma(s)&=
\frac{3}{4}\pi\sigma_0(s)\!\!\!
\sum_{q=u,d,s,c,b} \int d\hat{x}\;  dz dr dt \; \int dx_q dx_\qb\; 
\delta(\hat{x}-x_qx_\qb z t)
\\&\times
f^{h_1}_q(   s\hat{x}, x_q) 
f^{h_2}_\qb( s\hat{x}, x_\qb) \;
 \rho_I^{(0)}\big(\gamma_{Iq}(s\hat{x}/m_q^2),z\big)\; 
 \rho_I^{(2)}\big(-\gamma_{Iq}(Q_0^2/m_q^2),t\big)\; 
\\&\times
\sigma^{Born}_{q\qb}(s\hat{x}z)\;
\langle W_{MC} \rangle.
\label{eq:kkhhsigmaPru}
\end{split}
\end{equation}
We see that, as eq.(\ref{eq:kkhhsigmaPru}) includes an extra convolution with the well known second order exponentiated ISR
``radiator function'' $\rho_I^{(2)} $ with the negative evolution time argument
$-\gamma_{Iq}(Q_0^2/m_q^2)$ defined in Ref.~\cite{jad-yost-afb}, the QED below $Q_0$ is thus removed. This is illustrated in Fig.~\ref{fig4} from Ref.~\cite{ichep2022-say} 
for the $P_{T_\gamma}$ for the photon for which it is the largest in $Z\gamma^*$ production and decay to lepton pairs at the LHC at $8 ~TeV$ for
cuts as described in the figure. The results in the figure show that the effect of QED contamination in non-QED
PDFs is below the errors on the PDFs~\cite{kkmchh2}.\par
\begin{figure}[h]
\begin{center}
\setlength{\unitlength}{1in}
\begin{picture}(6.0,2.0)(0,0)
\put(0,0.5){\includegraphics[width=2.5in,height=1.5in]{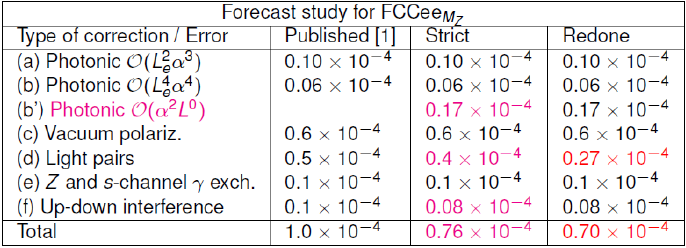}}
\put(2.55,0.5){\includegraphics[width=2.5in,height=1.5in]{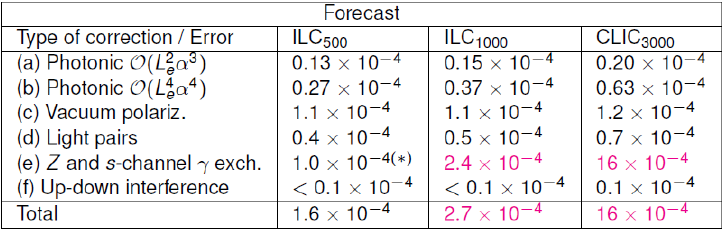}}
\put(1.3,0.25){(a)\hspace{2.3in}(b)}
\end{picture}
\end{center}
\vspace{-10mm}
\caption{\baselineskip=11pt Current purview on luminosity theory errors: (a), FCC-ee at $M_Z$; (b), proposed higher energy colliders}
\label{fig5}
\end{figure}

As another example of the gateway that YFS methodology provides for precision accelerator physics, we note that, for the planned EW/Higgs factories, we and our collaborators have discussed in Refs.~\cite{Jadach:2018jjo,Jadach:2021ayv-sh,fcc2023wkshpms} the new perspectives for the BHLUMI~\cite{bhlumi4:1996} luminosity theory error. In Fig.~\ref{fig5}~\cite{fcc2023wkshpms}, wherein we show the current purview for the FCC-ee at $M_Z$ and that for the proposed higher energy colliders,
this new perspective is exhibited. 
In addition to the improvements at $M_Z$ shown in Fig.~\ref{fig5}(a) to 0.007\%, there is the possibility that item (c) in Fig~\ref{fig5}(a) could be reduced by a factor of ~ 6 by the use of the results in Ref.~\cite{fjeger-fccwksp2019} together with lattice methods~\cite{latt1,latt2}. The formula to be studied is $$\Delta\alpha_{had}(t)=\Delta\alpha_{had}(-Q^2_0)|_{lat}+[\Delta\alpha_{had}(t)-\Delta\alpha_{had}(-Q^2_0)]|_{pQCDAdler}$$ with {\it lat} denoting the methods of Refs.~\cite{latt1,latt2} and {\it pQCDAdler} denoting the methods of Ref.~\cite{fjeger-fccwksp2019}. \par
Amplitude-based resummation applied to quantum gravity has been shown in Refs.~\cite{ward:2013dkunv,ijmpa2018} to tame its UV divergences. Using a standard notation, we note that one of the many consequences is  
\begin{equation}
\begin{split}
\rho_\Lambda(t_0)&\cong \frac{-M_{Pl}^4(1+c_{2,eff}k_{tr}^2/(360\pi M_{Pl}^2))^2}{64}\sum_j\frac{(-1)^Fn_j}{\rho_j^2}\cr
          &\qquad\quad \times \frac{t_{tr}^2}{t_{eq}^2} \times (\frac{t_{eq}^{2/3}}{t_0^{2/3}})^3\cr
    &\cong \frac{-M_{Pl}^2(1.0362)^2(-9.194\times 10^{-3})}{64}\frac{(25)^2}{t_0^2}\cr
   &\cong (2.4\times 10^{-3}eV)^4.
\end{split}
\label{eq-rho-expt}
\end{equation}
Here, $t_{tr}\sim 25 t_{Pl}$~\cite{reuter2,ward:2013dkunv,ijmpa2018} is the transition time between the Planck regime and the classical Friedmann-Robertson-Walker(FRW) regime and $t_0$ is the age of the universe and we take it to be $t_0\cong 13.7\times 10^9$ yrs.
In the estimate in (\ref{eq-rho-expt}), the first factor in the second line comes from the radiation dominated period from
$t_{tr}$ to $t_{eq}$ and the second factor
comes from the matter dominated period from $t_{eq}$ to $t_0$.
The experimental result~\cite{pdg2008}\footnote{See also Ref.~\cite{sola2,sola3} for analyses that suggest 
a value for $\rho_\Lambda(t_0)$ that is qualitatively similar to this experimental result.} 
$\rho_\Lambda(t_0)|_{\text{expt}}\cong ((2.37\pm 0.05)\times 10^{-3}eV)^4$ is close to the estimate in (\ref{eq-rho-expt}).
\par
\section{Gate to Precision Collider Physics: Improving the Collinear Limit in YFS Theory}
In the usual YFS theory the virtual infrared function $B$ in the s-channel resums (exponentiates)~\cite{Jadach:2023cl1} the non-infrared term
$\frac{1}{2}Q_e^2{\alpha\over\pi} L$ in $e^+(p_2)\;e^-(p_1)\rightarrow \bar{f}(p_4)\;e^-(p_3)$ using an obvious notation where the respective big log is $L = \ln(s/m_e^2)$ when $s=(p_1+p_2)^2$ is the center-of-mass energy squared. From Ref.~\cite{gribv-lptv:1972} we have that the term $\frac{3}{2}Q_e^2{\alpha\over\pi} L$ exponentiates -- see also Refs.~\cite{frixione-2019,bertone-2019,frixione-2021,bertone-2022} for recent developments in the
attendant collinear factorization approach. Does the YFS theory allow an extension that would also exponentiate the latter term and thereby enhance its role as a gate
to precision collider physics? This question has been answered in the affirmative in one of our last works with Staszek in collaboration with Prof. Zbigniew Was in Ref.~\cite{Jadach:2023cl1}.\par
Specifically, in Ref.~\cite{Jadach:2023cl1} we show that the virtual infrared function $B$ in the s-channel can be extended to
\begin{equation}
\begin{split}
B_{CL}&\equiv B+ \Delta{\bf  B}\\
          &= \int {d^4k\over k^2} {i\over (2\pi)^3} 
                       \bigg[\bigg( {2p-k \over 2kp-k^2} - {2q+k \over 2kq+k^2} \bigg)^2{\bf -\frac{4pk-4qk}{(2pk-k^2)(2qk+k^2)}}\bigg]
\end{split}
\end{equation}
and that the real infrared function $\tilde{B}$ can be extended to 
\begin{equation}
\begin{split}
\tilde{B}_{CL} &\equiv \tilde{B}+ \Delta{\tilde{\bf B}}\\
 &=\frac{-1}{8\pi^2}\int\frac{d^3k}{k_0}\bigg\{\large(\frac{p_1}{kp_1} - \frac{p_2}{kp_2}\large)^2 +{\bf \frac{1}{kp_1}\large(2 -\frac{kp_2}{p_1p_2}\large)}\\
                                          &\qquad\qquad+{\bf \frac{1}{kp_2}\large(2 -\frac{kp_1}{p_1p_2}\large)}\bigg\}.
\end{split}
\label{eq-real2}
\end{equation}
In an obvious notation, the extensions are indicated in boldface. While the YFS infrared algebra is unaffected by these extensions, the $B_{CL}$ does exponentiate the entire $\frac{3}{2}Q_e^2{\alpha\over\pi} L$ term and the respective collinear big log of the exact result in Ref.~\cite{berends-neerver-burgers:1988} in the soft regime is carried by the $\tilde{B}_{CL}$.\par 
For the CEEX soft eikonal amplitude factor defined in Ref.~\cite{ceex2:1999sh}
for the photon polarization $\sigma$ and $e^-$  helicity $\sigma'$ the corresponding collinear extension is given by 
\begin{equation}
\begin{split}
\sfac_{CL,\sigma}(k) = \sqrt{2}Q_ee\bigg[-\sqrt{\frac{p_1\zeta}{k\zeta}}\frac{<k\sigma|\hat{p}_1 -\sigma>}{2p_1k}
    +{\bf \delta_{\sigma'\;-\sigma}\sqrt{\frac{k\zeta}{p_1\zeta}}\frac{<k\sigma|\hat{p}_1 \sigma'>}{2p_1k}}\\
    + \sqrt{\frac{p_2\zeta}{k\zeta}}\frac{<k\sigma|\hat{p}_2 -\sigma>}{2p_2k}+{\bf \delta_{\sigma' \sigma}\sqrt{\frac{k\zeta}{p_2\zeta}}\frac{<\hat{p}_2 \sigma'|k -\sigma>}{2p_2k}}\bigg],
\end{split}
\label{eq-real4}
\end{equation}
where from Ref.~\cite{ceex2:1999sh} $\zeta\equiv (1,1,0,0)$ for our choice for the respective auxiliary vector in our Global Positioning of Spin (GPS)~\cite{gps:1998} spinor conventions with the consequent definition $\hat{p}= p - \zeta m^2/(2\zeta p)$
for any four vector $p$ with $p^2 = m^2.$ The collinear extension terms are again indicated in boldface.\par
These extended infrared functions are expected to give in general a higher precision for a given level of exactness~\cite{bflwetaltoappear}.\par
\section{Summary}
Prof. Staszek Jadach made seminal contributions to the role of YFS resummation in its role as a gate to precision accelerator physics experiments. His legacy in this area of theoretical physics is truly outstanding. He was my closest friend and dearest colleague. He was like a brother to me. As we can see from the discussion above, his legacy lives onward. We miss him dearly.\par
\bibliography{Tauola_interface_design}{}
\bibliographystyle{utphys_spires}

\end{document}